\documentclass[12pt]{article}
\usepackage{amssymb}
\usepackage{amsmath}
\usepackage{amsthm}

\begin{document}

\centerline{\Large \bf The Spin-Statistics Theorem in Arbitrary Dimensions}
\vskip 0.75cm

\centerline{Luis J. Boya\footnote{Permanent address: Departamento de
F\'{\i}sica
Te\'{o}rica. Universidad de Zaragoza. E-50009 ZARAGOZA,
Spain}(luisjo$@$unizar.es)}
\vskip 0.5cm
\centerline{and}
\vskip 0.5 cm
\centerline{E.C.G. Sudarshan (sudarshan$@$physics.utexas.edu)}
\vskip 0.5cm

\centerline{Center for Particle Theory, The University
of Texas at Austin.}
\medskip
\centerline{AUSTIN, TX 78712}
\medskip

\vskip 1cm

\begin{abstract}
We investigate the spin-statistics connection in arbitrary dimensions
for hermitian spinor or tensor quantum fields with a rotationally invariant
bilinear Lagrangian density. We use essentially the same simple method as
for space dimension $D=3$. We find the usual connection (tensors as bosons
and spinors as fermions) for $D=8n+3, \ 8n+4, \ 8n+5$, but only bosons
for spinors and tensors in dimensions $8n \pm 1$ and $8n$. In dimensions
$4n+2$ the spinors may be chosen as bosons or fermions.\\

The argument hinges on finding the identity representation of the
rotation group either on the symmetric or the antisymmetric part
of the square of the field representation.
\end{abstract}

\section{Introduction}

		The spin-statistics connection is an essential ingredient
in our description
of the world with quantized fields, which assures on one hand the existence of
macroscopic fields (like the radiation field), and on the other hand
gives rise through anticommuting fields (Pauli principle) to the valence
electrons, the chemical bonds etc., and therefore to the existence of
forms and structures: the Pauli principle is really the
{\it differentiating principle} in Nature.\\

	The connection asserts that the wavefunction of several identical
particles
in $D=3$ with integer spin remains unchanged under an arbitrary
permutation of the arguments, in which case bosons (Bose-Einstein
statistics, BE)
obtains, whereas for half-integer spin the permuted wavefunction picks up
a minus sign whenever one performs a transposition of the order
(or, more generally, an odd sign permutation) (Fermi-Dirac statistics, FD).
This translates in the usual way in the commutation  (BE) or
anticommutation (FD) of fields at different space points. \\

	With hindsight, we can say that historically the first case of such
correlation appeared in the statistical mechanics of {\it Lichtquanten}
or photons (from Planck 1900 to Einstein 1905, to Bose 1924).
But the main application came with the interpretation of the Pauli
exclusion principle (1925) by Heisenberg and by Dirac (1926) in terms
of the (anti-)symmetry of the many particle wavefunctions under
transposition, and subsequent application to many electron atoms.
This relation between spin and statistics is fundamental; it led
to the periodic system of chemical elements, peculiar intensity
rules in band spectra of homonuclear diatomic molecules, non-classical
scattering of alpha particles, or the Ehrenfest-Oppenheimer
theorem \cite{Ehr} (1931) on the statistics of compound systems,
among the oldest applications, and more recently to the selection
rules for the decay of unstable particles e.g. positronium,
coherent boson states and the existence of the laser, superfluidity
of helium four (Kapitza 1938) and later superconductivity as
coherent states of Cooper pairs (BCS theory, 1957) and even
superfluidity in helium three through condensation of pairs of
atoms.\footnote { The symmetry or antisymmetry of the wave functions
obtains only in a quantum field theory. We find remarkable
that a healthy positivistic attitude, namely that the permutation
of identical particles should produce no observable effects, would
have different ways to be implemented in the quantum theory!
This is possible at all, of course, because of the projective or
rather the bundle nature of quantum states, which we have already
emphasized \cite{Boya-ECG}. } \\

	There is more to the quantum states that mere rays. The wavefunction
is really a section on a vector bundle with base the space of rays
(projective space), and a sign change after permutation on the
arguments of the wavefunction is allowed as long as the associated
density matrix is invariant: in particular, a pure state lifted to a
vector repesentative could acquire a plus or minus sign,
corresponding to the two unique one-dimensional
 irreducible representations ({\it irreps}) of the symmetric
group; the sign cancels going down to the base space: in
this context, that is the precise form of the common
statement that quantum states are state vectors up to a
phase, here up to a sign. In mathematical terms, statistics
sign is just a Schur multiplier.\\

	The same argument shows also \cite{Boya-ECG} why half-integer
angular momentum could exist in quantum theory in the
first place; namely, the pertinent projective
representations of the rotation group (say, $SO(n)$) come
from the linear representations of the double covering
(universal covering for $ n > 2$), the $Spin(n)$ group. That is,
spin $1/2$ is an admissible (projective) representation of
the $3D$ rotation group, although it does not come from a
linear one, and picks up a minus sign under a full
rotation. \\

		We find it logical that (in space dimension $3$, and as
we shall see also in $D=8n+3, +4 \ \rm or \ +5$) these two
nonclassical properties "compensate each other";
namely, the case of Fermi statistics goes along with
 half-integer spin whereas the Bose statistics occurs with
integer spin. There are two compensating minus signs for
spinors (permutations of fermions {\it and} $2\pi $ rotation), but
none in the Bose (tensor) case, which therefore looks
more, but it is altogether not, wholly classical. This
connection is essentially the spin-statistics relation. \\

	 In this paper we are going to see whether the spin-statistics
 connection holds in arbitrary dimensions. The
motivation to study this question is fairly clear today:
unification of forces by the Kaluza-Klein mechanism,
supersymmetry and superstrings, extended objects and
$M-$theory, etc., all point to the necessity of higher
dimensions, whether invisible or macroscopic; in
$F-$theory we even face the case of (2, 10) spacetime
dimensions, that is, two times. As these theories are
quantum theories, one needs to see how the usual
argument, i.e. the symmetry of the bilinear scalar
product under $3D$ rotations, extends now to other $D>3$
dimensions. Although at the moment the question is
 rather academic, if one of these higher dimensional
theories stands in the future, the question will be an
important one, and so we believe that the present
investigation is justified. \\

\section{Review of the usual proof}
	There are many proof of the Spin-Statistics
relation in relativistic quantum field theory, starting with
the original one by Pauli in 1940 \cite{Pauli}; for a thorough
review of the situation up to the year 2000 see the book \cite{Sud-D}.
For our purposes we shall recall here the proof of the
theorem as given by Sudarshan many years ago \cite{Sud ECG},
which starts from a $3D$ rotationally invariant field
Lagrangian density and contains the essential features.
The manifold applications of the theorem in
nonrelativitic contexts claims for a demonstration not
requiring relativistic invariance. Axiomatic formulations
of quantum field theory, which do not use Lagrangians,
do need special relativity to prove commutativity
properties of the fields at distant points \cite{PCT}. However, the
requirement of relativistic invariance is somewhat
inappropiate, since most of the manifestations of this
relationship are in the nonrelativistic domain: atoms,
nuclei, condensed matter situations, quantum liquids,
phonons in solids, etc. Also the key topological feature,
namely the symmetry group not being simply connected,
appears already in the pure space part. \\

		The fundamental principle of field dynamics is the
Action Principle, as established by Weiss (1938) and in its
quantized form by Schwinger (1951) \cite{Schw}. This
presentation of the quantum theory demands that the
variation of any object $\Phi$ in the theory be given by its
commutator with the variation of $S$, the action of the
 system. That is
\begin{equation}
\delta \Phi = [\Phi, -i\delta S],
\end{equation}
which is simply the generalization of the quantum rule
$[q, p]= i\hbar $. It characterizes the action as universal
generator of variations. The action is the time integral of
the Lagrangian; we shall describe now a {\it classical}
mechanical theory in the first order formalism in which
the Lagrangian is a function on the $TT^{*}Q$ manifold, where
$Q$ (dim $Q=n$) is the configuration space, $T^{*}Q$ the phase
space (or cotangent bundle) and $TM$ means the tangent
bundle to any manifold $M$. In the first order formalism
we have the Lagrangian function $L_0 \in F(TT^{*}Q)$, with
\begin{equation}
L_0 = p_a \dot {q}_a - H(q_b, p_b)
\end{equation}
with summation on $a, 1 \leq a \leq n$, and where
 $\dot {A} = ( \partial / \partial t) A$.\\

	The equations of motion are not altered by the
(anti)symmetrization
\begin{equation}
L = 1/2(p_a \dot {q_a} - q_a \dot {p_a}) - H(q_b, p_b)
\end{equation}
	So, defining $ \xi := (q_a, p_a)$ as a column vector, we can
take the Lagrangian as ($^{t}B$ is the transpose of $B$)
\begin{equation}
L = 1/2( ^{t}\xi \,  C \, (\partial / \partial t)\xi) - H(\xi )
\end{equation}
where $C = -^{t}C$ is a purely numerical (invertible) real
antisymmetric matrix. Notice the "symplectic" character
of this first order Lagrangian associated to the use of first
order time derivative. The variable $\xi$ (with $2n$ number of
 components) depends in time, and the dynamical term
 $H(\xi)$ does not contain time derivatives. Notice also in this
formalism the kinetic term is bilinear in the fields.\\

			Inspired by this, we know write our Action and the
Lagrangian density operators for arbitrary {\it quantum}
fields $\chi $ as
\begin{equation}
S[\chi ]= \int _{t_0}^{t} dt \int d^{3}x {\cal L}[\chi],
 \, {\cal L } = (1/4) (^{t}\chi K (\partial / \partial t \chi) -
 (\partial / \partial t ^{t} \chi ) K \chi) - {\cal H} [\chi]
\end{equation}
following also Schwinger \cite{Schw} \cite{JS}. Here $ \chi = \chi ({\bf
x}, t)$ is a
finite-dimensional hermitian quantum field, ${\cal H}$ is the
Hamiltonian density, and $K$ is an {\it antihermitian}
numerical matrix: The dynamical variable becomes
hermitian, and $K$ should be taken antihermitian, $K=-K^{\dagger} $. \\

		But now there are naturally two possibilities: the
matrix $K$ can be real antisymmetric or purely imaginary
amd symmetric, as already Schwinger said half a century
 ago; and these two possibilities would fix the
commutation/anticommutation properties of the field
variation $ \delta \chi $ with the fields contained in $\chi $, leading
finally to the sought-for connection between spin and
statistics. \\

		Here we shall use for $\Phi $ just the fields $\chi
$. The
complete Lagrangian would have many pieces, viz.:
\begin{equation}
{\cal L} = {\cal L}_1 (kinetic, \, \,	 time) +  {\cal L}_2 (kinetic, \,
\,	 space)
+ {\cal L}_3 (mass \, \,) + {\cal L}_4 (\, \, interactions)
\end{equation}
where we know already that the first term, from
rotational invariance alone, should be a scalar and so we
shall impose $SO(3)$ invariance in the quantum
mechanical sense of above, that is, linear $SU(2)$
invariance. The form of the temporal kinetic term, with
the imposed rotational invariance, {\it is the only ingredient
we need for our proof of the theorem}. Whether the
remaining terms in the Lagrangian, specially the dynamics
encoded in the Hamiltonian, would spoil the arguments,
we leave open at this point and will comment later on. \\

		The general variation $\delta S$ contains three
terms: variation in the content of the integral inside the
fixed boundary, which gives the equations of motion,
variations of the limits of integration, and thirdly
variations of the field quantities at the fixed boundary.
For our case only the {\it third} variation is pertinent, namely
the variations of the fields at the boundaries, which can be
taken as two spacelike surfaces at times $t_0$ and $t$,
respectively: we consider the variation only on the
"future", at time $t$, and omit the (repeated) $t$ label. \\

		The equation becomes
\begin{equation}
4i\delta \chi_{a} (x) = [\chi_{a} (x), \int d^{3}y \, \{\delta \,
^{t}\chi_{b}(y)
K_{bc} \chi_{c}(y) - ^{t}\chi_{b} (y) K_{bc} \delta \chi_{c} (y) \}]
\end{equation}
where $x = ({\bf x},t), y =({\bf y}, t)$,etc. \\

	This is completely general. Now we require that \\

		" The field variation $ \delta \chi _{a} (x)$ either
conmmutes or
anticommutes with the field itself": this is equivalent to
restricting ourselves to fermi or bose statistics (we
specifically exclude parastatistics; the only kind of
parastatistics that is valid is the reducible parastatistics
as introduced by H. S. Green \cite{verde}; see also \cite{Fla}). \\

		A) The field variation COMMUTES with the field
itself. Then we obtain in the usual way (see \cite{Sud ECG} )
\begin{equation}
2i \delta ^{3} ({\bf x} - {\bf y}) = [\chi_{a} ({\bf x}), \chi_{b} ({\bf
y})]K_{ab}
\end{equation}
where $K$ has to be real {\it antisymmetric}. This matrix $K$
might be degenerate: call $K_{0}$ the restriction of $K$ to the
minimal components of a particular spin in $\chi $: $K_{0}$ is then
regular (invertible). We can then write symbolically
\begin{equation}
2iK_{0}^{-1} = [\chi, \chi]
\end{equation}
where
\begin{equation}
K_{0} = - ^{t}K_{0}  \  \   \rm with \  \  \  det \ K_{0} \neq 0
\end{equation}
which is the most general way of expressing the
fundamental commutation relations characteristic of
Bose fields: different field components commute, but
field and momentum components have an "i" as their
commutator. \\

		 B) The field variation ANTICOMMUTES with the field
itself. Then from the previous eq. we obtain, (with $ \{a, b\} :=ab +ba$):
\begin{equation}
2 \delta ^{3} ({\bf x} - {\bf y}) = \{ \chi_{a}({\bf x}), \chi_{b}({\bf
y}) \}K_{ab}
\end{equation}
with $K$ now a {\it real} symmetric matrix. Again, by restricting
to minimum fields, we can write the anticommutation
rules for Fermi fields in a form similar as before:
\begin{equation}
2 K_{0}^{-1} = \{ \chi, \chi \}.
\end{equation}
But the character of $K$ can be obtained also from the
kinetic term of the Lagrangian by appealing to rotational
invariance: namely $^{t}\chi K \partial /\partial t \chi $ has to be a
$SO(3)$ scalar
(invariant), as $K$ connects only pieces of $\chi$ with the same
spin. Recalling that the kinetic term involves the
{\it anti}symmetric time derivative, for {\it integer spin} the
matrix $K$ has to be {\it antisymmetric}, whereas for {\it halfinteger}
spin $K$ is {\it symmetric}:\ \

		In three space dimensions the squares of the
irreducible representations of $SU(2)$ are well known; for
example, for {\it l} integer
\begin{equation}
D_{{\it l}} \otimes  D_{{\it l}} = D_{0+} + D_{1-} + D_{2+} + ... +
D_{2{\it l}+}
\end{equation}
whereas for $s$ half-integer
\begin{equation}
D_{s} \otimes D_{s} = D_{0-} + D_{1+} + ... + D_{2s+}
\end{equation}
where $(+)$ indicates the symmetric, and $(-)$ the
antisymmetric, parts of the product. This says
that for tensors, the Identity {\it irrep} (scalar product) is in
the symmetric part, whereas for spinors is in the
antisymmetric part (e.g. $D_{1/2} \otimes D_{1/2} =D_{1} + D_{0} = 3(sym) +
1(asym)$). This crucial result comes really from the
{\it symplectic character} of the fundamental, spin $1/2$ {\it irrep}
of $SO(3)$, as $Spin(3)=SU(2) = SpU(1)$. \ \

		This encompasses the spin-statistics theorem in $3$
space dimensions: the specific form of $K = - K^{\dagger }$ in the
lagrangian implies that integer spin would have $K$
as real antisymmetric, hence the commutation relations
and Bose statistics. With half-integer spin fields is the
other way around: symmetric $K$ would imply
anticommutators, hence Fermi statistics and the Pauli
exclusion principle. \\

			The argument can be reversed, namely starting from
this result, we would conclude the symmetry/antisymmetry
of $K$, and from this the rules Bose or Fermi respectively
for BE or FD, recalling that the time derivative is an
antisymmetric operator. \\

				Our argument in $3D$ is really in
consonance with
special relativity: namely, the use of a spacelike surface
to state initial operator conditions is a wholly Lorentz
invariant statement. It is still valid for, say, Galilean
invariant theories, as long as one deals with quantum
field theories, in which particles can be created and
destroyed. \\

			We address now the question whether the validity of
the proof could not be spoiled by the neglected terms in
the Lagrangian. The space part of the kinetic density
should cause no problems, and indeed our argument
should be a proof of the spin-statistics theorem for
nonrelativistic {\it field} theories, in which particle
creation/destruction is allowed. Wightman has
emphasized that simple, quantum-mechanical many-body
systems with fixed number of particles need not obey the
standard spin-statistics relation. The reason for the
sufficiency of the time derivative comes from the
variation principle refering, in our case, to two spacelike
surfaces. \\

			What about limitations coming from peculiar
Hamiltonians? We do not have a full answer to this, but
would like to make the following remarks: in some cases, in
which the Hamiltonian is not bounded from below, as in
the naive case of the Dirac equation, the right statistics
comes to the rescue, and makes sense of such a
Hamiltonian, as the "anticommutator" statistics
incorporates the exclusion principle, and the equivalent
of hole theory and redefinition of the vacuum makes the
rest. A general Hamiltonian with no lower bound would
be of course unacceptable already at the classical level. \\

			Another question is the applicability of the method
to composite systems; it is a bit striking that e.g. protons
and neutrons, being fermions, make up compound
systems like the deuteron or the alpha particle with
tested boson character. Here we only remind the reader
of the old quoted Ehrenfest-Oppenheimer paper \cite{Ehr},
 making very plausible that composite even/odd number
of fermions should enjoy corresponding bose/fermi
statistics. But one should admit frankly that the whole
issue of statistics of composite systems deserves a closer
look. \\

\section{Particle statistics in arbitrary dimensions}
  Before going into technical details we would like to
show why the symmetry type of $K$ in $3D$ is {\it not} to be
expected for $8$ space dimensions. \\

	The reason is this: in $3D$, the Id {\it irrep} of $Spin(3)$
appears in the antisymmetric part of the square, viz.:
\begin{equation}
D_{1/2} \wedge D_{1/2} = D_{0} : \,  2 \otimes 2 = 3(sym) + 1(asym)
\end{equation}
whereas in $8D$ the two chiral irreps of $Spin(8)$, $\underline{8}_{s, c}$
behave like the vector irrep $\underline{8}_{v}$, also of dim $8$, because
triality (see e.g. \cite{car}) permutes the three {\it irreps}, so
the Id {\it irrep} appears necessarily in the symmetric square
of {\it any} of the three {\it irreps }:
\begin{equation}
\underline{8}_{v} \vee \underline{8}_{v} = \underline{1} + \underline{35}, \,
\underline{8}_{v} \wedge \underline{8}_{v} = \underline{28}
\end{equation}
\begin{equation}
\underline{8}_{s} \vee \underline{8}_{s} = \underline{1} +
\underline{35^\prime },
\, \,
\underline{8}_{s} \wedge \underline{8}_{s} = \underline{28} \, \,
\rm same \ \ \rm  for \ \  \underline{8}_{c};
\end{equation}
here $\underline{28}$ is the adjoint, $\underline{35}$ the $2-$symmetric
traceless,
$\underline{35^\prime }$ the (anti-)self-dual $4$-form, etc. Thus for $D=8$ the
identity
(Id) {\it irrep} appears in the symmetric part of the square of
either chiral {\it irrep}, contrary to the situation in $3D$; so
they can only describe bose fields, according the
arguments above. In the {\it Appendix} we delve more deeply
in the dimension $8$ case. \\

	Indeed, from the properties of Clifford algebra we
can see that the $8D$ case is a case of {\it real } type for the spin
{\it irreps}, whereas in $3D$ the type is {\it quaternionic}
(pseudoreal). The general result is now easily obtained
from the Clifford periodicity-$8$ theorem for spin groups,
which itself can be easily obtained from the {\it finite}
Clifford groups \cite{pajaro}. The result for the Type $T$ of $Spin(n)$
irrep is
\begin{equation}
Dim \, 8n+3, 8n+4, 8n+5 \, : T= -1 \, (peudoreal)
\end{equation}
\begin{equation}
Dim \, 4n+2, \, : T= 0  \, (complex)
\end{equation}
\begin{equation}
Dim \, 8n+1, 8n, 8n-1 \, : T= +1 \, (real)
\end{equation}

	In the first case, because $Spin(n)$ group lies inside
the symplectic part, the normal situation obtains. The Id
{\it irrep} is in the antisymmetric part. In the third case, is the
opposite: the Spin group lies in the orthogonal part, and
the Id {\it irrep} is in the symmetric part. This generalizes the
cases $D=3$ and $D=8$ respectively. \\

		In the complex case, $2$ mod $4$, the Id {\it irrep}, being
real, has to be in the (mixed) product of the two complex
conjugate {\it irreps}; by putting them together we get real
fields (Majorana). There are two Id {\it irreps}, and we can
always arrange to have one in the antisymmetric part, if we wish,
but it is not forced upon us. In other words, spinors in
$4n+2$ dimensions can be either bosons or fermions.\\

		Since the governing crtiterion is the Type, whether $R$
(real, $+1$), $C$ (complex, $0$) or $H$ (quasireal, -1), the general
result, as far as the argument depends on the group of
the space only, is:\\

		For $8n+3, \, +4, \, +5 \,$, the usual spin-statistics
connection obtains, and spinors are fermions.\\

	For $8n \pm 1, \, 0$, a wrong connection extants (i.e.,
tensors and spinors have to be bosons). \\

	For $4n+2$ (complex case), spinors can be fermions or
bosons, it is up to us. Tensors are bosons in all cases
(correspondence principle). \\

\section{Concluding remarks}
	We see that the proof of the commutation rules
in arbitrary dimension is very simple; it uses only the
temporal part if the kinetic term in the Lagrangian. This
is in the spirit of Neuenschwander query \cite{nuevo} regarding a
simple proof of the spin-statistics connection, extended
now to arbitrary dimensions. \\

		We find surprisingly few references to the $n-$dimensional
spin-statistics question in the published literature; the
reason might be that statistics deals with two or more
particles, but in higher dimensions we still have to find
one! Weinberg \cite{guru} is one of the few references to higher
dimension spin-statistics connection; see also \cite{reply}.\\

		The derived results are for wavefunctions of
theories with quantized fields, allowing variable number
of particles. For a {\it fixed} number of identical particles one
could use either symmetrized or antisymmetrized
wavefunctions \cite{Bac}. \\

	We are not considering dimensions $1$ and $2$. There is
no little group in $D=1$, as $SO(1)= \{1 \}$, hence no spin, and
indeed there is some freedom to chose the quantization
rules: recall quantized solitons in $1+1$ dimension should
behave as fermions (Coleman, Mandelstam 1975). Also,
the covering group of $SO(2)$ is $R$: then there is a vast
margin for statistics. The large literature for space
dimension $2$, where anyons live, han been reviewed e.g.
by Forte \cite{Forte}. \\

		\section{Appendix}
		We include here for completeness some
mathematical results regarding spin groups and spin
representations; see \cite{car} and \cite{pajaro} \\

		The first eitght spin groups already reflect the
{\it  Irrep} Type as stated above, because the isomorphisms \\
%\begin{equation}
%	Spin(1)=O(1), \, Type (+1) \, Spin(2)=U(1) = SO(2) \, Type (0) \\
%Spin(3) = SU(2) = SpU(1) \, Type (-1) \, Spin(4) = SpU(1)^{2},\, Type (-1) \\
%Spin(5) = SpU(2) \, Type(-1), \, \, Spin(6) = SU(4), \, Type (0)
%\end{equation}
\begin{equation}
\begin{array}{rcccccc}
& Spin(1) &Spin(2)&Spin(3)&Spin(4)&Spin(5)&Spin(6)\\
&||&||&||&||&||&||\\
&O(1)&U(1)&SpU(1)&SpU(1)^2&SpU(2)&SU(4)\\
Type&+1&0&-1&-1&-1&0
\end{array}
\end{equation}

and again $Spin(7)$ and $Spin(8)$ are real, Type (+1). \\

	In three dimensions, the spin group $Spin(3)$ has a
faithful {\it irrep} of complex dimension $2$, isomorphic
indeed to $SU(2) = SpU(1)$. The relation with the rotation
group
\begin{equation}
1 \rightarrow Z_2 \rightarrow SU(2) \rightarrow SO(3) \rightarrow 1
\end{equation}
implies spinors rotate $1/2$ turn when vectors make a full
turn. In $8$ space dimensions the situation is different. The
group $Spin(8)$ has as center $Z_2 \times Z_2$, so it has {\it three} order
two subgroups, whose quotients coincide with the two
real chiral spin representations $\underline{8}_s = \Delta_{L} \ , \
\underline {8}_c = \Delta_{R}$ and the vector one, $\underline {8}_{v}$.
A further quotient by the remaining $Z_{2}$ produces the projective group
$PO(8)$ from either:
%\begin{array}
%\end{array}
\begin{equation}
Spin(8)\left\{
\begin{array}{ll}
Spinor_L&=\Delta_L,\dim\, 8 \quad real\\
Vector &=SO(8)\\
Spinor_R&=\Delta_R,\dim\,8 \quad real
\end{array}
\right\}\quad PO(8)
\end{equation}

		Now in $8$ dimensions the three real dim$-8$
representations are permuted by the symmetric group in
three symbols $S_{3}$ (Cartan's triality, \cite{car}). Recall $O(8)$,
with symbol $D_{4}$ is the {\it unique} simple Lie algebra with a
large than $Z_{2}$ automorphism group. Therefore the square
of either {\it irrep} of the three should be similar, so it is
impossible that the chiral {\it irrep} and the vector {\it irreps}
differ in the symmetry type of the product. Indeed, the
products of these $8-$dim irreps are (with $+$ sym, $-$ asym parts)
$$\begin{array}{cl}
Vector^2&=graviton (35+) + dilaton (1+)+{\rm 2-form} (28-)\\
Spinor_L^2&=selfdual+{\rm 4-}form (35+)+scalar (1+)+{\rm 2-}form(28-)\\
Spinor_R^2&=antiselfdual \ {\rm 4-}form (35+)+scalar (1+)+{\rm 2-}form (28-)\\
Vector \times Spinor&= Gravitino (56)+vector (8)\\
Spinor_L\times Spinor_R&={\rm 3-}form (56)+vector(8)
\end{array}
$$
	We use the particle content of SuperGravity ${\cal N}=2$ in
ten dimensions, whose massless little group is $O(8)$. \\

	Indeed, the exceptional Lie algebra $F_{4}$ contains the
four fundamental {\it irreps} of  $D_{4}$: dim $F_{4} = 52 =
8\times 7/2 + 3\times 8$. The Weyl group of $F_{4}$ (order 1152) is the
symmetry group
of the $24-$cell, the most symmetric of the regular polytopes,
living in $4$ dimensions \cite{Cox}, and tessellating $S^{3}$. \\


\begin{thebibliography}{AAAAA}

\bibitem{Ehr}
Ehrenfest, P. and Oppenheimer, J.R., Phys. Rev. {\bf 37}, 333 (1931).

\bibitem{Boya-ECG}
Boya, L.J. and Sudarshan, E.C.G., Found. Phys. Lett. {\bf 4}, 283 (1991).

\bibitem{Pauli}
Pauli, W., Phys. Rev. {\bf 58}, 716 (1940). Reprinted in
J. Schwinger, {\it Selected Q.E.D. papers}. Dover 1956

\bibitem{Sud-D}
Duck, I., and Sudarshan, E.C.G., {\it Pauli and the
Spin-Statistics Theorem}. World Scientific 1997.


\bibitem{Sud ECG}
Sudarshan, E.C.G., Proc. Ind. Acad. Sci. {\bf 67}, 284 (1968).
Id. in Proc. Nobel Symposium 8, Almquist
Wiksell 1968.- Shaji, A., and Sudarshan, E.C.G.,
ArXiv: quant-ph/0306033.


\bibitem{PCT}
Strater, R. and Wightman, A.S., {\it PCT, Spin-Statistics
and all That}. Benjamin 1964.


\bibitem{Schw}
Schwinger, J., Phys. Rev. {\bf 82}, 914 (1951). Reprinted
in \cite {Pauli}. See also his book {\it Quantum Kinematics and
Dynamics}, Addison-Wesley 1991 (orig. ed. 1970).

\bibitem{JS}
Schwinger, J., Proc. Nat. Acad. Sci. USA {\bf 44}, 228 and 619 (1958).

\bibitem{verde}
Green, H.S. Phys. Rev. {\bf 90}, 279 (1953)

\bibitem{Fla}
Flaherty, F., in {\it Directions in general relativity}, Vol.2.-
B.H. Lu ed., Cambridge U.P. 1993, p. 125

\bibitem{car}
Cartan, E. {\it La Theorie de Spineurs}, Vols. 1 and 2.
Hermann, Paris 1938

\bibitem{pajaro}
Boya, L.J. and Byrd, M., J. Phys. {\bf A 32}, L201 (1999)

\bibitem{nuevo}
Neuenschwander, D.E., Am. J. Phys. {\bf 62}, 972 (1994)


\bibitem{guru}
Weinberg, S. Phys. Lett. {\bf B 143}, 97 (1984)

\bibitem{reply}
Ahluwalia, D.V. and Ernst, D. J.,
Phys. Rev. {\bf C 45}, 3010 (1992)

\bibitem{Bac}
Bacry, H., Am. J. Phys. {\bf 63}, 297 (1995)

\bibitem{Forte}
Forte, S., Rev. Mod. Phys. {\bf 64}, 193 (1992)

\bibitem{Cox}
Coxeter, H., {\it Regular Polytopes}. Dover 1973


\end{thebibliography}
\end{document}